\documentclass[10pt,conference]{IEEEtran} 
\IEEEoverridecommandlockouts

\usepackage{url} 
\usepackage{framed}
\usepackage{mdframed}
\usepackage{listings}
\usepackage{multicol}
\usepackage{amssymb}
\usepackage{lipsum}
\usepackage{adjustbox}
\usepackage[font=bf,skip=\baselineskip]{caption}
\usepackage{tabularx}
\usepackage{seqsplit}
\usepackage{multirow}
\usepackage{booktabs}
\usepackage{graphicx}
\usepackage{hyperref}
\usepackage{subcaption}
\usepackage{makecell}
\usepackage[table]{xcolor}
\usepackage[ruled,vlined,linesnumbered]{algorithm2e}
\usepackage{amsmath}
\usepackage{tikz}
\usepackage{tcolorbox}
\usepackage{threeparttable}

\usepackage{breakurl} 
\usepackage{boldline}
\usepackage[ruled,vlined]{algorithm2e}

\usepackage{pifont}
\usepackage{balance}
\let\labelindent\relax
\usepackage{enumitem}
\usepackage[htt]{hyphenat}
\usepackage{array}

\definecolor{customblue}{HTML}{006ca6}
\definecolor{customgreen}{HTML}{009264}
\definecolor{custombrown}{HTML}{ff3d00}
\AtEndPreamble{
\usepackage{hyperref}

    \hypersetup{
      colorlinks = true,
      linkcolor = customblue,
      anchorcolor = purple,
      citecolor = customblue,
      filecolor = purple,
      urlcolor = customblue
    }
}

\newcommand{\tool}{\textsc{CommitShield}}

\begin{document}

\title{\tool{}: Tracking Vulnerability Introduction and Fix in Version Control Systems}

\author{
\IEEEauthorblockN{Zhaonan Wu\IEEEauthorrefmark{2}, Yanjie Zhao\IEEEauthorrefmark{2}\IEEEauthorrefmark{1}, Chen Wei\IEEEauthorrefmark{3}\IEEEauthorrefmark{1}, Zirui Wan\IEEEauthorrefmark{2}, Yue Liu\IEEEauthorrefmark{4}, Haoyu Wang\IEEEauthorrefmark{2}
}

\IEEEauthorblockA{\IEEEauthorrefmark{2}Huazhong University of Science and Technology, Wuhan, China\\
wuzhaonan680@gmail.com,  yanjie\_zhao@hust.edu.cn, ziruiwan@hust.edu.cn, haoyuwang@hust.edu.cn}
\IEEEauthorblockA{\IEEEauthorrefmark{3}MYbank, Ant Group, Hangzhou, China, 
juyi.wc@mybank.cn}
\IEEEauthorblockA{\IEEEauthorrefmark{4}Monash University, Melbourne, Australia, yue.liu1@monash.edu}

\thanks{\IEEEauthorrefmark{1}Yanjie Zhao (yanjie\_zhao@hust.edu.cn) and Chen Wei (juyi.wc@mybank.cn) are the corresponding authors.}
\thanks{\IEEEauthorrefmark{2}The full name of the author's affiliation is Hubei Key Laboratory of Distributed System Security, Hubei Engineering Research Center on Big Data Security, School of Cyber Science and Engineering, Huazhong University of Science and Technology.}
}
\maketitle

\begin{abstract}
Version control systems are commonly used to manage open-source software, in which each commit may introduce new vulnerabilities or fix existing ones. Researchers have developed various tools for detecting vulnerabilities in code commits, but their performance is limited by factors such as neglecting descriptive data and challenges in accurately identifying vulnerability introductions. To overcome these limitations, we propose \tool{}, which combines the code analysis capabilities of static analysis tools with the natural language and code understanding capabilities of large language models (LLMs) to enhance the accuracy of vulnerability introduction and fix detection by generating precise descriptions and obtaining rich patch contexts. We evaluate \tool{} using the newly constructed vulnerability fix dataset, CommitVulFix, and a cleaned vulnerability introduction dataset. Experimental results indicate that \tool{} improves recall by 74\%-77\% over state-of-the-art methods in the vulnerability fix detection task, and its F1-score improves by 15\%-27\% in the vulnerability introduction detection task.

\end{abstract}

\section{Introduction}
\label{sec:introduction}
Version control systems~\cite{islam2018sentiment} play an irreplaceable role in maintaining and managing open-source software (OSS). The \texttt{commit}, as a core concept within the version control system, records every change made in the software code. With the widespread application of OSS in software development, vulnerabilities in OSS can pose a serious threat to the software system. In a version control system, when a vulnerability is addressed, the fix information is updated through a \texttt{commit}. The National Vulnerability Database (NVD)~\cite{nvd2024}, established by the National Institute of Standards and Technology (NIST), contains extensive information about software vulnerabilities. This often includes links to known affected software and fixes for related versions, with the \texttt{commit} link being a crucial reference for vulnerability fixes.
 
Disclosure of vulnerabilities in OSS follows the Coordinated Vulnerability Disclosure (CVD) model. Details of a vulnerability are disclosed only after the developer confirms sufficient time has passed for a fix to be implemented. However, the time between submitting a fix and disclosing the vulnerability is often not immediate, creating a window for potential malicious exploitation. Therefore, before formal disclosure, developers can conduct vulnerability fix detection on commits to gather specific information about undisclosed vulnerabilities. This information is essential for remediating vulnerabilities in open-source components, thereby maintaining software security during development.
Furthermore, analyzing the commits that introduced vulnerabilities in the version control system can enhance the collection of vulnerability information. In certain cases, if a software project uses a version that contains an introduced but unrepaired vulnerability, the system may be exposed to security threats due to the vulnerability-related code. In such situations, developers can use the commit information related to the introduction of the vulnerability to guide their repair efforts according to system requirements. Therefore, designing an automated tool for detecting both the introduction and fixes of vulnerabilities is of significant importance. Such a tool can enhance the security features of version control systems and mitigate the impact of vulnerabilities during the development process.

Vulnerability fix detection (VFD) and vulnerability introduction detection (VID) are challenging tasks. In the realm of VFD, researchers have developed several automated tools. For example, VulFixMiner~\cite{zhou2021finding} uses the advanced pre-trained model CodeBERT~\cite{feng2020codebert} to analyze code changes but neglects the commit descriptions, thereby missing important contextual information. VulCurator~\cite{nguyen2022vulcurator} improves upon VulFixMiner by incorporating commit descriptions into its analysis, assessing problem descriptions, and ultimately presenting results as probabilities. However, these tools focus primarily on code changes, lacking a comprehensive examination of patches in conjunction with their descriptions and failing to gather contextual information related to the patch code. This limitation leads to a high number of false negatives.
In terms of VID, the SZZ algorithm~\cite{sliwerski2005changes} is commonly referenced. Originally designed to detect the introduction of common errors, various improved SZZ algorithms have been proposed, including B-SZZ~\cite{sliwerski2005changes}, AG-SZZ~\cite{kim2006automatic}, RA-SZZ~\cite{neto2018impact}, and V-SZZ~\cite{bao2022v}. V-SZZ is the latest iteration that employs a line mapping algorithm to identify the earliest commits associated with the modified lines of code related to vulnerability introduction. Similar to earlier algorithms, V-SZZ is effective in identifying patches that fix vulnerabilities by removing code, but lacks the ability to identify patches that fix vulnerabilities by adding code.

Large Language Models (LLMs) have demonstrated remarkable capabilities in natural language processing and code understanding~\cite{hou2023large, nam2024using}, making their application to vulnerability introduction and fix detection promising. To address the limitations of existing tools, we propose \tool{}, an approach that leverages static analysis tools and LLMs to enhance vulnerability introduction and fix detection.
For VFD, \tool{} processes key commit information in two steps. First, it generates a detailed description of the commit. Second, it uses static analysis tools to obtain more comprehensive context related to the patch, which is then provided to the LLM for analysis and result generation.
In terms of VID, \tool{} first identifies the lines of code modified by the vulnerability fix patch and gathers additional context to locate the file containing the vulnerability. It then retrieves the patches from the historical commits of that file and supplies the collected information to the LLM for analysis, resulting in outputs for further detection. 

To summarize, the paper makes the following contributions:
\begin{itemize}
    \item We propose a novel tool, \tool{}, that detects vulnerability introductions and fixes in commits by combining static analysis tools and LLMs.
    \item We collect and release a new dataset called CommitVulFix for the evaluation of VFD task, comprising 681 C/C++ vulnerability fix commits and 1,118 non-vulnerability fix commits since 2023. For VID, we utilized and cleaned the dataset proposed in V-SZZ, resulting in a dataset containing 284 vulnerability introductions.
    \item Experimental results show that \tool{} consistently outperforms state-of-the-art (SOTA) methods in VFD, with recall improvements of 74\% to 77\%. We also established a baseline that does not provide any additional information to the LLM. Under this baseline, the precision of \tool{} improved by 19\%, and the F1-score increased by 13\%. For VID, \tool{} generally outperforms existing SZZ algorithms, with an F1-score improvement of 15\% to 27\%. Furthermore, we conducted a case study to highlight samples that other tools failed to detect, demonstrating \tool{}'s effectiveness in identifying real cases. 
\end{itemize}

\noindent\textbf{Artifact Accessibility.} The replication artifact is available at \url{https://github.com/security-pride/CommitShield}.
\section{Background and Motivation}
\label{sec:background}

\subsection{Patch Commit}
\begin{figure}[ht!]
    \center
    \includegraphics[width=0.8\linewidth]{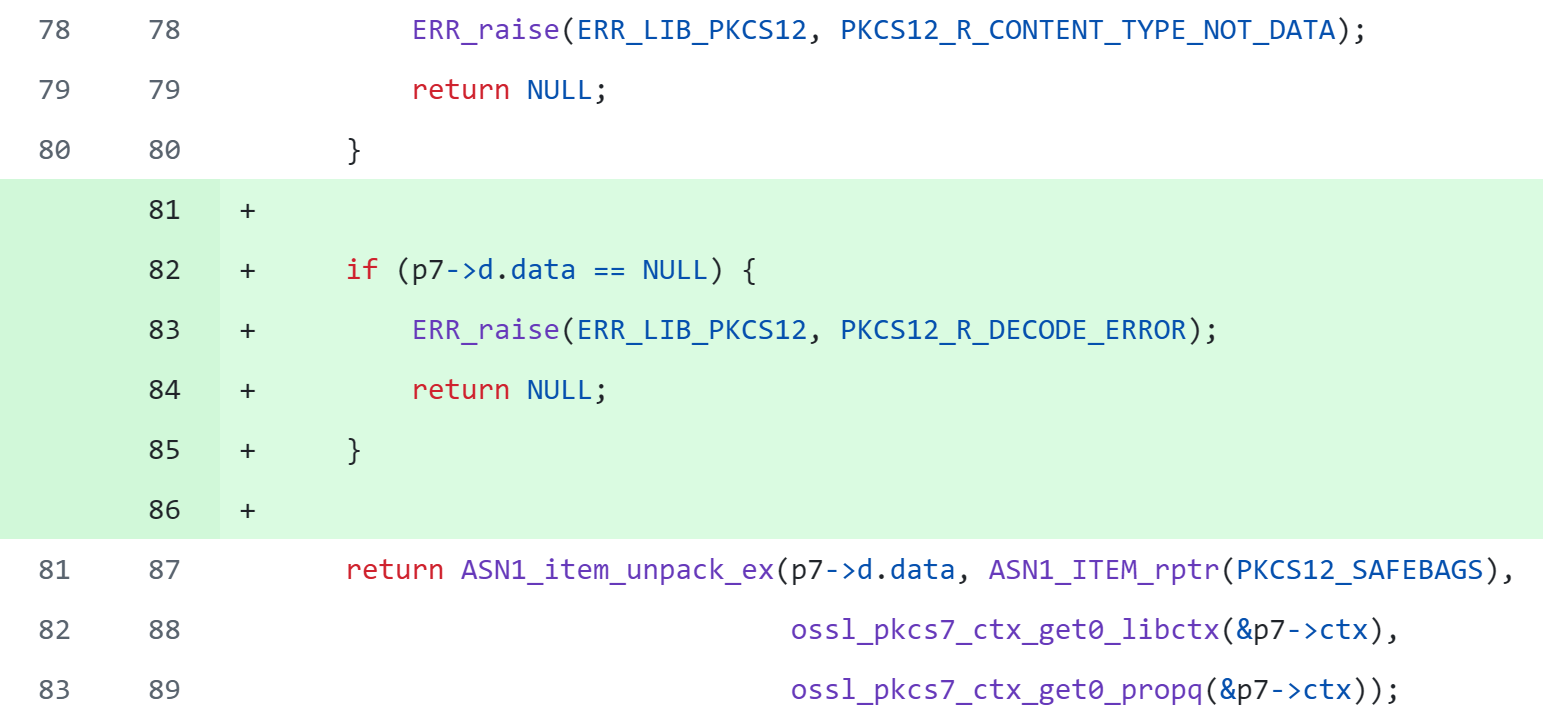}  
    \caption{Example of secure commit patch.}
    \label{fig:background}
\end{figure}

Patch commits, referred to as patches, generally consist of a code change and a description of the change~\cite{sawadogo2022sspcatcher}, with patches focusing specifically on modifications to code updates (e.g., the introduction of new functionality). Nowadays, source code patch commits play an increasingly important role in all phases of the software development life cycle. A commit typically consists of a commit message and a source code diff (also known as a “diff”), which is a record of changes made to the code between different software releases. In software maintenance, a single update to the codebase usually consists of many patch commits. Some of these patches are not related to security issues, such as feature upgrades and performance improvements; the commits that are used to fix security issues are called security commits~\cite{wang2021patchrnn}, which tend to receive more attention because they prevent vulnerabilities from being used maliciously to further jeopardize the entire system. \autoref{fig:background} illustrates a security commit that resolves a security issue by adding lines 82 to 85 to address a vulnerability that could cause a crash due to improper handling of NULL values in these structures. Existing work in the area related to patch commits is presented in~\cite{zuo2024vulnerability}, which mainly includes dataset construction~\cite{reis2021ground, wang2021patchdb}, different approaches to patch commit mining~\cite{perl2015vccfinder,tan2021locating, wang2023graphspd, zhou2023tmvdpatch}, and research in commit information~\cite{islam2018sentiment, zuo2023commit}.

\subsection{Vulnerability Introduction and Fix Detection} 
The VFD task takes a patch commit as input and identifies whether the patch commit has completed a vulnerability fix or not by analyzing code changes and change descriptions. This technique plays an important role in contemporary software version management systems. Software version-based VFD was first proposed in~\cite{tian2012identifying}, where vulnerability fix patches are identified in their scheme by code difference feature extraction and bag-of-words representation of commit messages. In~\cite{wang2019detecting}, code differences became the sole object of study; 61 feature sets were used to form vectors in machine learning, and an algorithm integrating five classifiers was designed to improve the precision of vulnerability fix patch identification. In recent years, semantic learning-based approaches for vulnerability patch detection have been proposed. \cite{zhou2021finding} proposed an approach called VulFixMiner, which focuses on differential code changes and uses CodeBert~\cite{feng2020codebert} to enhance VFD. In~\cite{nguyen2022vulcurator}, the VulCurator tool acquires richer information and uses deep learning to improve the F1-score in VFD.

The VID task takes as input a single patch commit of a vulnerability fix and detects, in a systematic way, in which historical patch commits the vulnerabilities were introduced. In~\cite{woo2021v0finder}, the authors refer to the origin of vulnerable software as Vulnerability Zero (VZ) and design a tool called V0Finder to accurately uncover VZs. In~\cite{bao2022v}, the authors upgrade the SZZ algorithm by proposing the V-SZZ algorithm, which further backtracks on early changes in vulnerable code to identify vulnerability-introducing commits.

\subsection{Motivation}
\label{sec:motivation}
The main challenges in VFD based on patch submissions are as follows.
\textbf{(1) Neglect of patch description.} Patch description, as an important part of a patch commit, significantly impacts the identification of vulnerability fix patches. However, deep learning-based tools~\cite{tian2012identifying, wang2019detecting} tend to focus solely on the code change part of the patch commit in their design, resulting in a false-positive rate as high as 41.3\% in~\cite{wang2019detecting}. This is a key factor limiting the performance improvement of such tools. Among PLM-based VFD tools, VulFixMiner also neglects description information. The VulCurator pays attention to description information, but due to the uneven quality of patch descriptions, its effectiveness also will encounter challenges. For example, the description of vulnerability CVE-2023-4683 is simply ``fixed \#2563'', and its detailed information appears in the issue related to the patch commit.
\textbf{(2) Diversity of patch code modifications.} In real software version control systems, the number of security patch submissions and non-security patch submissions tends to be disparate. Most of the patch submissions have nothing to do with vulnerability fixes, and non-security patch submissions may contain feature updates, code upgrades, performance enhancements, formatting adjustments, etc. These variations may still result in a high number of false positives and false negatives when tools attempt to identify vulnerability patches~\cite{zuo2024vulnerability}. This can lead to the neglect of real vulnerability fix patches and an over-concern with non-vulnerability fix patches, greatly reducing the efficiency of researchers in VFD.

A primary challenge in VID arises from analyzing patch submissions, i.e.,
\textbf{incomplete types of vulnerability introduction submissions}. 
The traditional VID approach, such as the SZZ algorithm and its V-SZZ variant, assumes that the code removed during the vulnerability fixing process is often the code where the vulnerability was originally introduced.
However, this assumption overlooks cases where a vulnerability introduction can instead be identified by analyzing newly added code.
Similarly, the V0Finder method requires the presence of deleted lines of code in the vulnerability patch code. The incomplete vulnerability-introduced commit type overlooks vulnerabilities that are fixed by adding lines of code. While looking for the location of vulnerability introduction through code that has been removed in a patch is the majority of cases, this does not mean that detecting the introduction of a vulnerability by analyzing the code that has been added is something that can be ignored. 
\section{Methodology}
\label{sec:methodology}

\begin{figure*}[ht!]  
    \centering     
    \includegraphics[width=\textwidth]{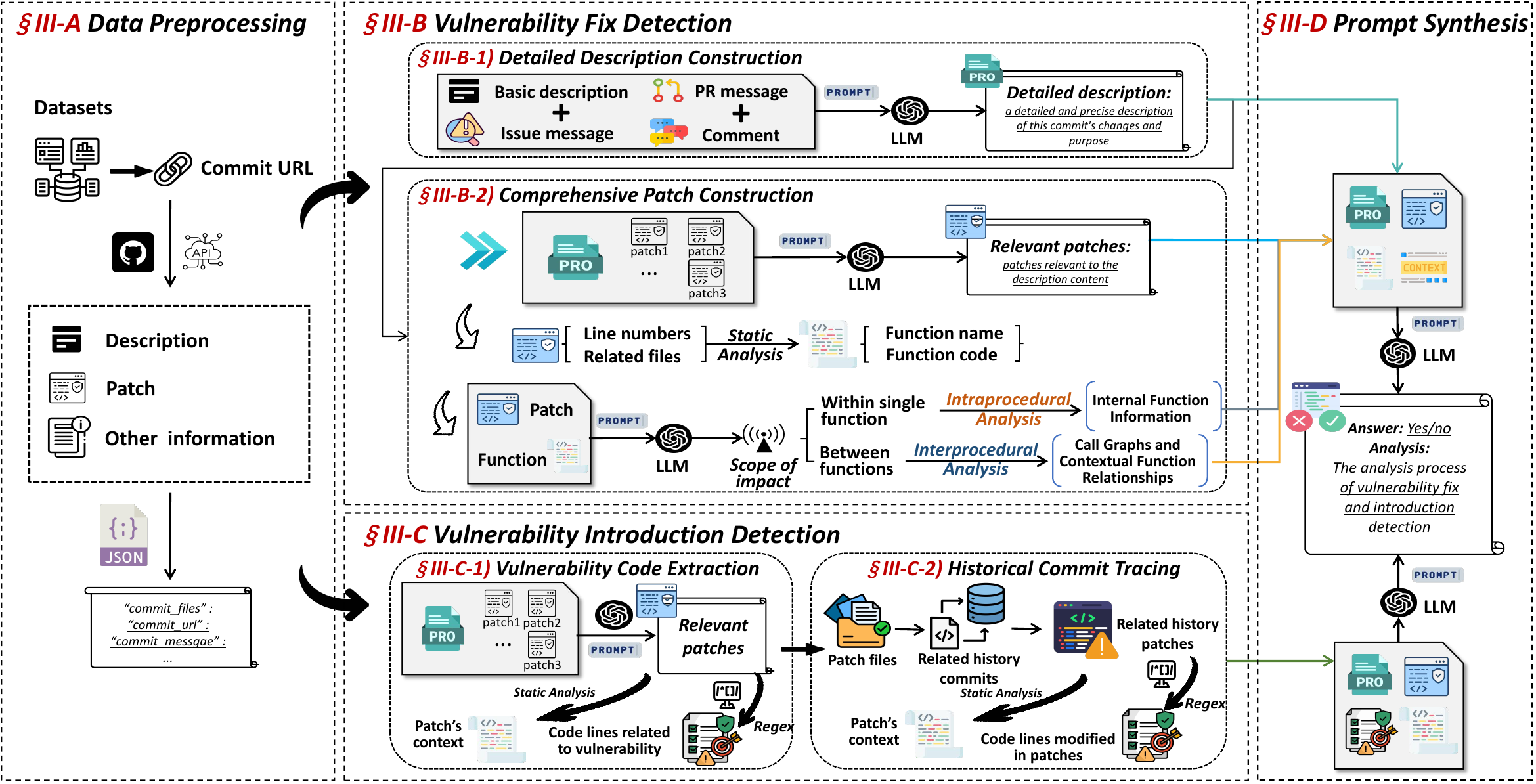}  
    \caption{Overview of \tool{}.}  
    \label{fig:framework}  
\end{figure*}

In this paper, we present \tool{}, an automated approach designed to address both VID and VFD tasks.
The overview of \tool{} is illustrated in \autoref{fig:framework}.

\subsection{Data Preprocessing}
\label{sec:data preprocessing}

For \tool{}, the expected input is a link to a commit for a particular repository. After obtaining that commit link, the system interacts by calling GitHub's API~\cite{github_api} to retrieve the details of that commit in the specific repository. In addition to recording the description and patch of the commit, we also extract other key information from the commit for usage in subsequent processes, including the commit URL, parent-commit URL, etc. Finally, we store this information in a dictionary in JSON format, allowing the system to retrieve the required information at each step according to the different tasks to be performed.

\subsection{Vulnerability fix detection}
\label{sec:fix detection}

\subsubsection{Detailed Description Construction}
\label{sec:description construction}
After examining a multitude of commits, we observe that relying solely on the commit description and code changes to identify vulnerability fixes is challenging, especially when the descriptive information is of low quality. This challenge is exemplified by the case of CVE-2023-4683 (as discussed in \autoref{sec:motivation}), in which more detailed commit information is only available within the associated issue. Drawing inspiration from this, when the base description includes an issue-related link (commonly formatted as ``\#number''), we can utilize GitHub's API to fetch the issue details as part of the commit description. In addition to issues, another prevalent type of link in the commit description points to pull requests. Since issue and pull request numbers are unique, we can uniformly access these links within the base description to retrieve specific information related to the commit. Furthermore, we find that some commit comments directly relate to the patch and can indicate whether the commit addresses a vulnerability. To encapsulate our approach, we collect the following information: the base commit description, issue details, pull request information, and relevant comments. We then construct a prompt incorporating these four types of information and leverage the advanced natural language processing capabilities of LLMs to generate a more precise and detailed description for the commit. If a commit lacks descriptive information beyond the base description, LLM will expand upon the basic description to produce a clearer natural language interpretation. The description generated in this step will be utilized for subsequent correlation analysis.

\subsubsection{Comprehensive Patch Construction}
The patch of a commit may contain various types of modifications, some of which may be irrelevant to the description. We categorize vulnerability fix-related modifications as security modifications. While some modifications aim to improve code functionality, focus on enhancing system performance, and still others pertain to comments, which we term as non-security modifications. During the VFD process, if a patch includes both secure and non-secure modifications, the non-secure ones can introduce noise that interferes with the identification of vulnerability fixes. Notably, modifications that enhance a security feature rather than repair an existing vulnerability are particularly prone to being misidentified as security modifications, thereby significantly impacting the vulnerability detection outcomes. To mitigate the impact of non-security modifications in VFD, we construct prompts for each patch described in the message and commit generated in \autoref{sec:description construction}, and utilize the natural language processing and code understanding capabilities of LLM to identify patches related to the description message. This step filters out the noisy patches in the commit, sorting and retaining only patches relevant to the description.

Based on the number of functions involved in a vulnerability, we classify them into two categories: intra-procedural and inter-procedural vulnerabilities~\cite{li2024effectiveness}. Intra-procedural vulnerabilities are typically confined to a single function, and their impact is limited to that function. In contrast, inter-procedural vulnerabilities involve call relationships between different functions, allowing their impact to propagate across functions that may reside within a single file or span multiple files.
In a commit-based VFD task, the input patch usually contains the modified lines of code along with their surrounding context. When the patched vulnerability is intra-procedural, it can be challenging to identify the fix based solely on the description and patch, especially if the number of modified lines is small and functional context is lacking. Even with manual verification, ensuring the accuracy of this type of VFD can be difficult. Similarly, when a patch addresses an inter-procedural vulnerability, the actual trigger statement for the vulnerability may not be included in the patch. Relying solely on the description and patch to identify this type of vulnerability is also complex.

To address these challenges, we have integrated a component into our framework to assess the scope of a patch's influence. 
If the impact of a code modification within a patch is confined to a single function, we consider the scope of the patch to be that function. Conversely, if the modification could potentially affect other functions, we assign the scope of the patch to those functions, pending further confirmation.
We begin by extracting the line number(s) of the modified line(s) and the patch file(s) from the relevant patch, followed by retrieving the complete code of file(s) from the parent commit version. We then utilize a static analysis tool, Tree-sitter\footnote{\url{https://tree-sitter.github.io/tree-sitter/}}, to analyze the line number(s) and code in the file(s), extracting the complete code of the function(s) that contains the modified line(s). We then use regular expressions to capture the full name of every function, preparing for potential future function call(s) analysis.
In particular, some modified line(s) may relate to header file addition, deletion, or macro definition change, which do not belong to any specific function. We handle these cases separately to ensure they do not interfere with the Tree-sitter analysis process. Once this step is complete, we obtain both the function code and the name of the function in which the modified line(s) resides within the patch.

After obtaining the patch and identifying the function(s) in which it is located, the next challenge is to analyze the scope of the patch's impact. We construct a prompt for the patch and its associated function(s), and then leverage the analysis capabilities of LLM to determine the role of the patch within that function. The LLM analyzes the impact of the modified line(s) in the patch and subsequently examines the context in which line(s) is affected through the provided function(s). The analysis results are subsequently generated; if the patch's impact is confined to the function, we consider the impact to be intra-procedural. We retain the patch and function data for the final VFD analysis. If the patch's impact may extend beyond the function, we proceed to perform a function call analysis on the functions where the patch is located. Specifically, we first download the patch's repository using \texttt{git} commands and switch the repository to the parent commit version. We then utilize Joern\footnote{\url{https://github.com/joernio/joern}} to generate a Code Property Graph (CPG)~\cite{yamaguchi2014modeling} for the current repository, extracting the call relationships of the functions under investigation in this version, including the file location of the call points, function information, and line details. Finally, we extract the context surrounding the call points from these call relationships and construct a collection of function call contexts related to this patch.

\subsubsection{Summary}
Having gathered all relevant information regarding VFD, we now summarize it as follows. For the description in the patch commit, we gather additional information related to the description to compile as much information about the commit as possible. We then utilize LLM to analyze this information and generate a more detailed commit description. For the patches within the commit, we first conduct a relevance analysis to retain the patches associated with the description information. Subsequently, we categorize the patches into intra-procedural and inter-procedural patches based on their impact scope. For the intra-procedural patches, we prepare information about the functions in which they are located. For the inter-procedural patches, we gather context regarding the function calls related to these patches. We organize this information for the final VFD analysis.

\label{sec:prompt synthesis}

\begin{figure*}[ht!]  
    \centering     
    \includegraphics[width=\textwidth]{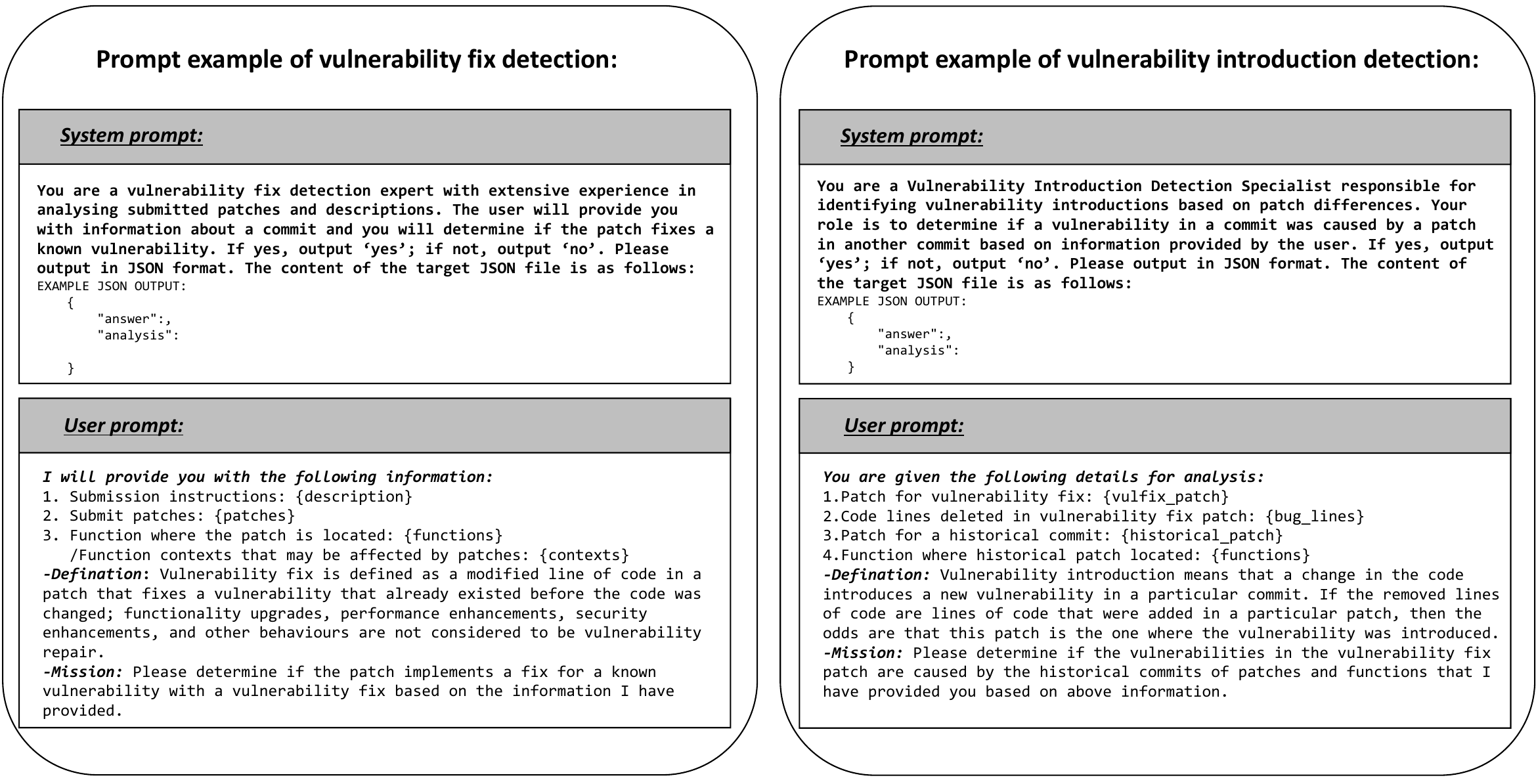}  
    \caption{The prompt example of VFD and VID mission.}  
    \label{fig:prompt}  
\end{figure*}

\subsection{Vulnerability Introduction Detection}
\label{sec:fix detection}

\subsubsection{Vulnerability Code Extraction}
For the VID task, we rely primarily on descriptive information and code modifications within the patch as the main body of analysis. A vulnerability introduction is defined as the point in the code submission history where the vulnerability was first introduced. The first step we take is to extract the code related to the vulnerability fix in the patch. By analyzing a large number of vulnerability fix patches, we have found that the vast majority of vulnerability fixes involve the deletion of a line of code. Even when the intention is to modify a part of a line of code, the line is typically presented in the patch as a deletion followed by an addition. If a vulnerability fix patch does not contain deleted code but only adds code to achieve the fix, then the added code is usually closely related to its context. Therefore, we deem it crucial to perform a comprehensive extraction of code modification lines in vulnerability fix patches. In \tool{}, we first filter out patches related to vulnerability fixes based on the vulnerability fix commit description. We then extract the deleted and added code lines from these patches using regular expressions and store the code lines in a dictionary. When the context in a patch is limited, we expand the context of the patch in order not to miss critical statements related to the vulnerability, and this expansion is capped at the full function where the patch is located. In addition to this, we also consider the types of vulnerabilities introduced as inter-procedural type which typically involves modifications to certain key variables or fields, creating vulnerabilities because they are not updated globally. For this type of potential inter-procedural vulnerability introduction, we utilize Tree-sitter to analyze the patch and identify the key variables within the modified lines. Subsequently, we employ \texttt{git} commands to review the historical changes to those variables, and we record the commit history related to these modifications, including the associated files and line numbers.

\subsubsection{Historical Commit Tracing}
To identify the historical introduction location of a vulnerability, tracing and analyzing the commits prior to the vulnerability fix commit is essential. Initially, we gather information about the vulnerability fixing commit, including the repository and the patch file. Subsequently, we employ the git tool to download the repository and revert it to the version at the time of the vulnerability fixing commit, retrieving all historical commits that modified the patch file up to that version. To ensure that the patch information in the history commits is as complete as possible, we extend the context of each patch.  We collect the start and end lines of patches and set extension boundaries for them, capped at the complete function in which the patch is located. This is with the exception of modifications that are not internal to the function, for which we measure the length of the patch. We define the number of lines in a patch as \( x \). When \( x \) is less than 10, we extend the patch by \( x \) lines both forward and backward; when \( x \) is greater than 10 but less than 30, we extend the patch by \( \frac{x}{2} \) lines both forward and backward; and when \( x \) is greater than 30, we consider that there is sufficient context information, rendering further extension unnecessary.

\subsubsection{Summary} 

We now summarize our \tool{} approach for VID. During the code extraction phase, we acknowledge that both deleted and added lines of code can influence vulnerability resolution. Therefore, we extract all types of patch modification lines to aid in subsequent VID. Besides, to ensure that no critical context is overlooked, we obtain an enriched context for the patch. In the history commit traceability phase, we not only review the fundamental patch contents from historical commits but also expand the context of these patches. 

After analyzing 50 vulnerability introductions, it was observed that software version control systems frequently record instances of vulnerability remediation. As mentioned in~\cite{rodriguez2020bugs}, typically, when a commit introduces a vulnerability, the immediately following commit often contains modifications intended to address the vulnerability. Therefore, during the phase of result output, \tool{} identifies and marks commits that do not yield any output results, signifying the absence of identification for the commit where the vulnerability was introduced within historical commits. Building upon our preceding analysis, in instances where no such commit is identified, \tool{} is employed to designate the most recent historical commit as the potential vulnerability introduction commit. Although this empirical method may occasionally result in false positives, it offers an approach to identify actual instances of vulnerability introductions.

\subsection{Prompt Synthesis}
Having collected all the necessary information for the two detection tasks, we now construct our prompts for the LLM. Prompt engineering~\cite{white2023prompt} facilitates interaction with the LLM and modulates its responses. The selection of prompt cues is crucial to the analytical outcomes of the LLM; therefore, precisely articulated and formatted cues are essential for enhancing the LLM's ability to analyze specific issues. It is noteworthy that the length of LLM prompts is subject to constraints—in our experiments (as detailed in \autoref{sec:evaluation}), the LLM prompt is limited to 130,000 tokens. If the prompt length exceeds this threshold, we must abbreviate the prompts to prevent system failures due to exceeding the limit.

The construction of the LLM prompt is illustrated in \autoref{fig:prompt}. The initial segment includes the data we present to the LLM, which consists of the commit patch and the generated information necessary for addressing a specific task. The subsequent segment outlines the task directives provided to the LLM, detailing the definitions of vulnerability fixes and introductions, and clarifying the tasks the LLM is expected to perform. Ultimately, we instruct the LLM to generate the required information, encompassing the outcomes of its analysis (either affirmative or negative) and the analytical process, with the final results presented in JSON format.

\subsection{Implementation}
\label{sec:Implemention}

In \tool{}, we selected the appropriate tool based on its required functionality, ensuring that it meets the criteria of high performance and user-friendliness. During the information acquisition phase, we utilized the GitHub REST API to gather detailed information about each commit. In the code analysis phase, we employed the Tree-sitter to ascertain the function in which the specified patch resides, capitalizing on its efficiency in extracting nodes of a particular type. Joern's proficiency in generating code attribute graphs led us to select it as our static analysis tool for determining function call relationships. When considering LLMs for analysis, we evaluated leading LLMs based on factors such as ease of use and cost-effectiveness. Ultimately, we opted for Deep-Seek-V2.5~\cite{zhu2024deepseek} as the LLM for analysis. Although there exist more powerful LLMs, their closed-source nature and high cost don't meet the user-friendliness requirements of \tool{}.

\section{Evaluation}
\label{sec:evaluation}

In \autoref{sec:dataset}, we present the methodology for constructing and selecting the datasets utilized for evaluation. Moving on to \autoref{sec:evaluation VFD}, we evaluate the effectiveness of \tool{} in addressing the VFD task. Subsequently, in \autoref{sec:evaluation VID}, we analyze the performance of \tool{} on the VID task. Finally, in \autoref{sec:case study}, we explore instances where \tool{} successfully identifies vulnerabilities that state-of-the-art (SOTA) methods fail to detect accurately.

\subsection{Dataset Construction}
\label{sec:dataset}

\subsubsection{Dataset of VFD}
In the VFD task, we are not only concerned with the system's ability to accurately identify genuine vulnerability fix commits, but we are also focused on its capability to correctly identify non-vulnerability fix commits. Frequent misidentification of non-vulnerability-fixing commits by the system can impose a significant additional workload on security personnel. Therefore, it is essential to have a comprehensive dataset that includes both vulnerability fix commits and non-vulnerability fix commits to evaluate the performance of SOTA and \tool{} in the VFD task. Given that C/C++ is widely regarded as the most vulnerable programming language~\cite{ wang2019detecting, wang2021patchrnn, wang2020machine}, we have selected the commits from C/C++ repositories as our detection target.

Although extensive work has been conducted on vulnerability data collection, and these efforts have made the collected data publicly available on the Internet, the data often lacks the most recent vulnerabilities~\cite{fan2020ac, wang2021patchdb, zheng2021d2a}. Considering that LLM has the characteristic of using a large amount of data for training, there is a risk of data leakage if a known dataset is employed for evaluation. Consequently, we elected to collect C/C++ vulnerabilities from January 2023 to the present to form a dataset of vulnerability fixes. We adhered to the vulnerability collection methodology outlined in~\cite{wang2024reposvul} to gather data on C/C++ vulnerabilities disclosed from January 2023, retaining only those vulnerabilities that include a link to the relevant GitHub commit, indicating that the vulnerability has been fixed. Subsequently, we cleaned the collected vulnerability fix data, discarding those that did not contain C/C++ code changes in the patches. This process culminated in a vulnerability fix dataset comprising 681 C/C++ vulnerabilities.

Given that the number of non-vulnerability-fixing commits significantly exceeds that of vulnerability-fixing commits, we analyzed the repositories containing these vulnerabilities and selected 11 of them as the source for extracting non-vulnerability-fixing commits. Before initiating formal collection, we analyzed the descriptions of 681 C/C++ vulnerabilities to identify and summarize the set of keywords characteristic of vulnerability fix commits. Following this, we downloaded the 11 selected repositories and retrieved all commits from January 2024 to the present. We set four criteria to filter out non-vulnerability fix commits: first, the commit description should not include vulnerability fix keywords; second, the patch should involve changes to C/C++ code; third, the patch should affect fewer than three files; and fourth, the commit should not appear in the vulnerability fix dataset which we have already collected. It is important to note that the third rule is designed to ensure that the length of the input prompt does not surpass the token limit of LLM, thereby maintaining the system's validity. In the end, we collected 1,118 non-vulnerability fix data and randomly selected 100 of them for manual verification. The verification confirmed that all 100 items were indeed non-vulnerability fix commits, which validates the accuracy of our data.

Finally, we obtained the validation dataset for VFD, which we named CommitVulFix. This dataset comprises 681 C/C++ vulnerability fix commits and 1,118 C/C++ non-vulnerability fix commits.

\subsubsection{Dataset of VID}
To facilitate the comparison of the performance between \tool{} and SOTA, we selected the vulnerability introduction dataset collected in V-SZZ as our benchmark dataset, on which V-SZZ demonstrates superior performance compared to other SZZ algorithms. Before utilizing this dataset, we first conducted a cleaning process, removing inaccessible data. Consequently, we obtained a dataset comprising 284 vulnerability introductions.

\subsection{The effectiveness of \tool{} in VFD}
\label{sec:evaluation VFD}

To evaluate the effectiveness of \tool{} in VFD, we selected two VFD tools based on the pre-trained model CodeBert implementation: VulFixMiner and VulCurator. VulFixMiner focuses solely on code changes, while VulCurator extends VulFixMiner's capabilities by incorporating the analysis of descriptive information. We successfully deployed these two VFD tools as referenced in~\cite{nguyen2022vulcurator, zhou2021finding}. Although these two tools are not fine-tuned solely with C/C++ languages, they utilize the pre-trained model CodeBERT, which has the capability to extend applications to other programming languages. Additionally, we define the method of constructing a prompt that uses only the base commit information as a baseline, which has no additional information and we use the same LLM, Deep-Seek-V2.5, to perform the VFD task. 

We validated the performance of these tools and baselines on the constructed dataset CommitVulFix. We selected precision, recall, and F1-score as the evaluation metrics for effectiveness. Precision reflects the tools' ability to identify non-vulnerability-fixing commits, while recall reflects their ability to identify vulnerability-fixing commits. The F1-score indicates the tools' combined ability to identify these two types of commits. We recorded the results of VFD in \autoref{tab:rq1}. It is evident that \tool{} outperforms both the baseline and other VFD tools in terms of precision, recall, and F1-score. Compared to the baseline, \tool{} achieved a 19\% increase in precision and a 13\% increase in F1-score. We observed that the baseline's recall was as high as 94\%. Despite collecting the latest vulnerability-fixing data for evaluation, we suspect that the high recall rate suggests the possibility of data leakage in vulnerability-fixing data. Nonetheless, \tool{} detected 13 more vulnerability fixes than the baseline. This has significant implications for improving the identification of vulnerability fixes in version control systems. When compared to VulFixMiner and VulCurator, \tool{} improved precision by 19\%-23\%, recall by 74\%-77\%, and F1-score by 56\%-59\%. These results reflect the performance improvement of \tool{} over SOTA in accurately identifying vulnerability fix commits.

One possible factor for the improved performance of our tool compared to SOTA methods is that Deep-Seek has more parameters than CodeBERT, which might contribute to better performance. To address this concern, we selected a subset of the data for manual analysis. Results show that VulFixMiner does not utilize commit descriptions, and although VulCurator increases its focus on descriptions, its classifier-based prediction approach struggles when descriptions are unclear and patch code is simple. As demonstrated in the VFD case in \autoref{sec:case study VFD}, detailed vulnerability descriptions and more adequate context are necessary to maximize the capability of VFD. The lower precision of the baseline compared to \tool{} also indicates that the performance gains in improving the accuracy of identifying non-vulnerability fixes are not solely due to the increased number of parameters in the LLM.

\begin{table}[h]
\caption{Performance metrics across different VFD approaches.}
\begin{tabularx}{\linewidth}{l*{4}{>{\centering\arraybackslash}X}}
\hline
\textbf{Approach} & \textbf{Model} & \textbf{Precision} & \textbf{Recall} & \textbf{F1-score} \\ \hline
Baseline            & Deep-Seek                  & 0.62          & 0.94          & 0.75 \\
VulFixMiner         & CodeBERT                   & 0.58          & 0.22          & 0.32 \\
VulCurator          & CodeBERT                   & 0.62          & 0.19          & 0.29 \\
\tool{}             & Deep-Seek                  & 0.81          & 0.96          & 0.88 \\
\hline
\end{tabularx}
\label{tab:rq1}
\end{table}

\subsection{The effectiveness of \tool{} in VID}
\label{sec:evaluation VID}

To evaluate the effectiveness of \tool{} in VID, we selected the latest V-SZZ algorithm and various previous versions of the SZZ algorithm for comparison with \tool{} evaluation. The validation dataset we utilized is the vulnerability introduction dataset provided in V-SZZ, which contains a total of 284 C/C++ vulnerability introductions after our cleaning process. Before delving into the evaluation results, we first elucidate the performance metrics for VID. In previous analyses of vulnerability introduction performance~\cite{bao2022v}, the recall has been the most representative metric, as it visually reflects the effectiveness of detecting actual vulnerability introductions. Precision is another significant metric; a higher precision indicates that the system introduces fewer false positives while identifying real vulnerability introductions. F1-score is the combination of the above two metrics. Furthermore, through our observations and findings, the commits associated with vulnerability introduction may not be unique. In this dataset, a vulnerability is tagged with a corresponding vulnerability introduction commit. This leads us to consider that when the detection result includes multiple vulnerability introduction commits, the dataset may identify other commits that could lead to vulnerability introduction as false positives, potentially affecting the evaluation results of the system. To facilitate comparison with existing SZZ algorithms, we disregard the impact of this situation and present the evaluation results of \tool{} on the VID task. As shown in \autoref{tab:rq2}, \tool{} outperforms existing SZZ algorithms in both precision and recall. Specifically, precision improved by 19\%-31\%, and recall improved by 3\%-35\% over other SZZ algorithms. Additionally, we calculated the F1-score for each tool, and \tool{}'s F1-score improved by 15\%-27\% compared to other SZZ algorithms. This demonstrates \tool{}'s high accuracy in identifying vulnerability introduction submissions, minimizing the number of false positives while accurately pinpointing the location of vulnerability introductions.

\begin{figure*}[ht!]  
    \centering     
    \includegraphics[width=\textwidth]{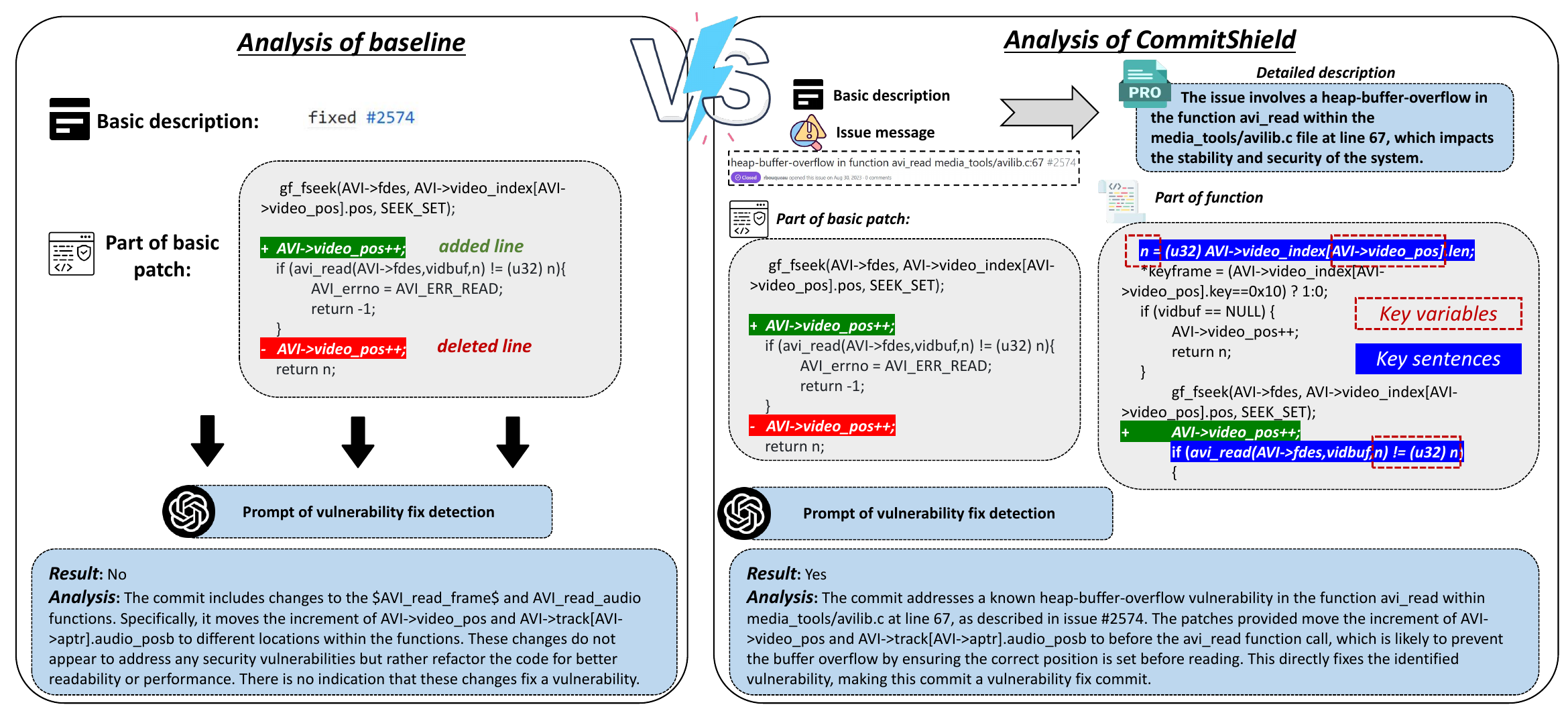}  
    \caption{Case study of VFD.}  
    \label{fig:case1}  
\end{figure*}

\begin{table}[h]
\centering
\caption{Performance metrics across different VID approaches.}
\begin{tabularx}{\linewidth}{l*{4}{>{\centering\arraybackslash}X}}
\hline
\textbf{Approach} & \textbf{Precision} & \textbf{Recall} & \textbf{F1-score} \\ 
\hline
V-SZZ            & 0.52         & 0.79          & 0.63 \\ 
AG-SZZ           & 0.49         & 0.63          & 0.55 \\ 
B-SZZ            & 0.46         & 0.67          & 0.55 \\ 
L-SZZ            & 0.55         & 0.47          & 0.51 \\ 
MA-SZZ           & 0.43         & 0.63          & 0.51 \\ 
R-SZZ            & 0.46         & 0.67          & 0.55 \\ 
\tool{}          & 0.74         & 0.82          & 0.78 \\ 
\hline
\end{tabularx}

\label{tab:rq2}
\end{table}

\subsection{Case study}
\label{sec:case study}

In order to visually present the ability of \tool{} to identify real vulnerability fixes and introductions, we have prepared two cases for readers to learn. 

\subsubsection{Case study of VFD}
\label{sec:case study VFD}
The first case pertains to VFD. We selected a case that was not identified by either SOTA or the baseline for analysis, comparing the analysis process and results of both the baseline and \tool{}. The CVE number for this case is CVE-2023-4682. In the baseline process depicted in \autoref{fig:case1}, we constructed the prompt using the basic description and patch. The commit description reads "fixed \#2574", which contains no additional commit information beyond its issue link. The basic patch indicates a location change in the sentence \texttt{AVI-\textgreater{}video\_pos++}. Due to space constraints, we present only part of the patch in \autoref{fig:case1}; the other part of the patch has the same function as the modification in \autoref{fig:case1}. Ultimately, the LLM provided its output and analysis, and it is evident that the LLM cannot identify the vulnerability fix commit solely based on its basic description and patch.

It can be seen that the baseline failed to recognize the vulnerability and instead interpreted the commit as performing code refactoring. Now, we demonstrate the process of \tool{}. We first visited the link in the description and found the issue related to this commit. The information in the issue indicates a buffer overflow in another function. Therefore, after \tool{}'s collection of other relevant descriptions, the LLM generated a detailed description.

\begin{figure*}[]  
    \centering     
    \includegraphics[width=\linewidth]{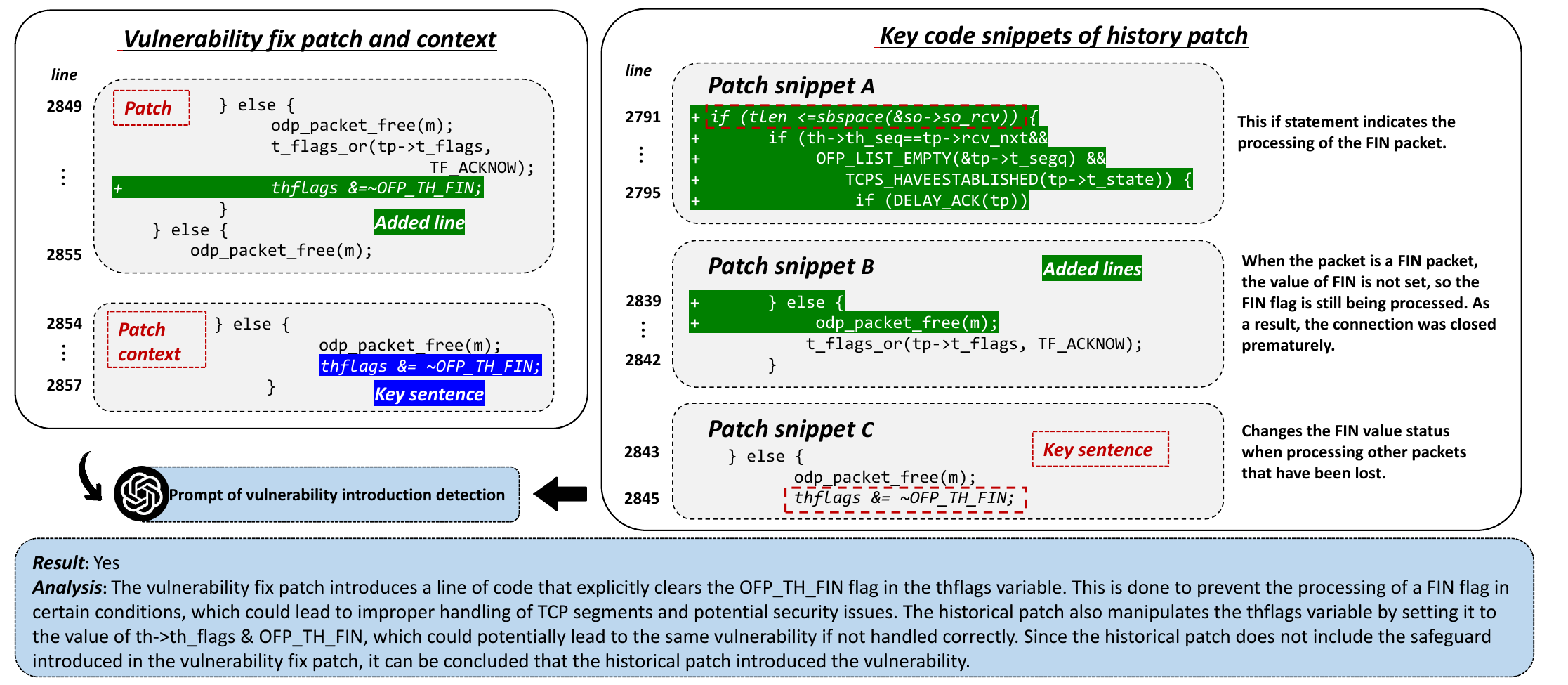}  
    \caption{Case study of VID.}  
    \label{fig:case2}  
\end{figure*}

Subsequently, we analyze the function where the patch is located and observe that the key variable \( n \), which is related to the modification statement, is not assigned within the patch. However, it appears both in the parameter of the  \texttt{avi\_read} function and in the judgment statement. When a heap buffer overflow occurs, \texttt{avi\_read} will return an error message, and if the value of \texttt{AVI->video\_pos} is not updated, then the value of \( n \) will not change, and at this point it will keep causing the heap buffer overflow, and will not move on to the next frame to be processed. At this point \tool{}can get richer context information by getting the function where patch is located, including the relationship between \( n \) and \texttt{AVI->video\_pos} and combine it with the detailed description for analysis. The final analysis by \tool{} is shown in \autoref{fig:case1}. Unlike the baseline result, \tool{} successfully identifies this vulnerability fix commit. 

\subsubsection{Case study of VID}

The second case pertains to VID. The vulnerability fix patch, as shown in \autoref{fig:case2}, addresses a known vulnerability by adding lines of code. The patch clears the FIN flag of the FIN packet, thereby resolving the issue where the FIN packet, when dropped, still has its FIN flag processed, causing the connection to close prematurely. As mentioned in our motivation, existing SZZ algorithms focus more on vulnerability fixes that involve deleted code. Therefore, we selected a vulnerability introduction case that none of the historical SZZ algorithms have successfully identified.

For general vulnerability introductions, deleted lines of code often contain vulnerability-related information. For vulnerability introductions implemented by adding lines of code, the context of the location where they are added usually contains information about the vulnerability (e.g., similar code snippets). Therefore, \tool{} extracts the contextual code snippets for such patches and traces back to where that part of the code snippet appeared. Eventually, \tool{} identified the commit patch shown in \autoref{fig:case2}. The first patch snippet, where the If statement is located, indicates the determination of whether there is enough \texttt{so-rcv-sockbuf} space when processing FIN packets. The second patch snippet triggers the retransmission mechanism for FIN packets. The third patch snippet is the processing of retransmission for other packets, which includes clearing the FIN flag. Therefore, by analyzing the semantics of the code in the patch and the code changes in the historical patches, \tool{} concludes that the added sentence in the vulnerability fix patch is functionally similar to patch snippet C in the historical patch. Consequently, the commit is marked as a vulnerability-introducing commit, and its analysis is output.

\section{Discussion}
\label{sec:discussion}

\subsection{Failure Analysis}
\label{sec:rq4}

\subsubsection{Failure Analysis of VFD}
In VFD, we analyze the false negative and false positive cases generated during the evaluation process. Among the 24 false-negative cases, we found that there are two types of patches that account for a relatively large number of these cases. The first type involves the addition of new files or header files. When the committed patch contains these additions, the vulnerability fix patch often involve new function calls. Our approach switches the snapshot of the repository to the parent commit version during the commit switching step, which will not produce any results for function call analysis. Consequently, \tool{} can only complete the test based on the generated description information and the original patch, and the test result is always negative. The second category involves modifications related to precise variables that are not defined by the local function. When \tool{} analyzes this type of commit, the patch may not belong to any of the functions, and therefore the relevant function calls are not available. At this point, \tool{} is still only able to analyze based on the generated description and the underlying patch, and the final analysis is always negative. Although the percentage of these two types of patches in the false negatives is relatively high, the recall of \tool{} in VFD is ultimately as high as 96\%, which still demonstrates the effectiveness in VFD.

In the false-positive cases, we find that the function implemented by the largest percentage of patches is usually the prevention of potential vulnerabilities. Changes to such patches do not fix vulnerabilities that already exist in the code, but rather avoid the creation of new vulnerabilities by enhancing the robustness of the code and improving its security defenses. Although this type of code does not fix any vulnerabilities, its implementation is closely related to security. Thus, even though \tool{} generates some false positives when detecting such changes, these false positives, if related to security prevention, will be of great benefit in improving the robustness of the software. In our dataset, the precision of \tool{} is as high as 81\%, and manual detection of this portion of false positives is not too labor-intensive. In conclusion, although \tool{} cannot identify all non-vulnerability fixes perfectly, for most commits, it is able to generate fewer false positives with accurate identification.

\subsubsection{Failure Analysis of VID}
In VID, we also analyze the false negative and false positive cases generated during evaluation. In the false-negative cases, we identified the following types that lead to false-negatives: the patch is unreadable, the file name is modified, and the traceback is ended prematurely after a false-positive data appears. The first kind of false negative is due to the diff file in the patch being too large to retrieve patch information using GitHub REST API; the second kind of false negative occurs because, in the process of backtracking the history of a specified file, a commit makes a change to the filename, which makes it impossible to backtrack to the real vulnerability-introducing commit; the third kind of false negative is due to the vulnerability introduction commit being too far away from the vulnerability fixing commit, resulting in the traceback ending within ten commits after a false positive occurs, and therefore the real vulnerability introduction commit is missed. A real example is the case of the 9 false positives before a real vulnerability introduction commit was detected.

In the case of false positives, we found that there are two main components that lead to false positives. One is that there may be some false positives that occur before the real vulnerability introduction is detected. These false positives usually have a high overlap in code fragments with the real vulnerability introduction commit and are sensitive to the changed lines in the vulnerability fix patch that are not vulnerability fixes, leading to parts of the commit unrelated to vulnerability fixes being flagged as false positives. The other part is when there is neither a true positive nor a false positive output, we mark the closest commit as a vulnerability introduction commit. Although the number of false positives is increased, this number constitutes a small percentage, and the verification can be done with less labor.

\subsection{Threats to Validity}
\label{sec:Threats}

\subsubsection{Internal Validity}
In evaluating the performance of VFD, we employed keyword matching to construct the non-vulnerability fix dataset and all commits have fewer than four files.. Although 100 pieces of data were randomly selected for analysis, we discovered the presence of some vulnerability fix data through failure analysis. These vulnerability fix data lack CVE IDs, yet they function as vulnerability fixes. In evaluating the performance of VID, we utilized the vulnerability introduction dataset constructed by our predecessors. In this dataset, one vulnerability corresponds to only one introduced commit, thus the presence of multiple commits introducing a vulnerability is not accounted for, and a fraction of true positives are mistakenly identified as false positives.

\subsubsection{External Validity}
Since the LLM is a key component of \tool{}, the final performance of \tool{} is closely linked to the performance of the LLM. Besides, we combine the static analysis tools Tree-sitter and Joern to obtain additional context information for patches. Since these tools are designed to analyze modifications within functions, they will fail when the modifications are not located within a function. In such cases, \tool{} can only rely on the generated description and basic patches for analysis, which may affect the detection results.

\section{Related Work}
\label{sec:related work}
\textbf{\textit{Patch commits and vulnerabilities.}} Some methods rely on commits in version control systems to address vulnerability-related issues, such as VFD and VID. In a recent study, GraphSPD~\cite{wang2023graphspd} developed a security patch system named PatchCPG, which utilizes a novel graph structure PatchCPG to represent the patch. In a recent study, \cite{zhou2023tmvdpatch} extracted control and data dependence graphs to gather semantic and syntactic details of code changes. They then used a BiLSTM model to encode commit messages and the paths derived from these decomposed graphs. \cite{le2024latent} conducted an empirical study on the potential and impact of using latent vulnerabilities in software vulnerability prediction, finding that these undocumented latent vulnerabilities before vulnerability fix commits could significantly enhance the performance of vulnerability prediction models. \cite{jiang2024understanding} represents the first empirical investigation of kernel vulnerability introduction commits (KVICs) in the Linux kernel. A total of 1,240 KVICs were identified, and their characteristics, purposes, and associated human factors were analyzed. This research provides insights for improving vulnerability detection and prevention in the development of the Linux kernel. Unlike these approaches, \tool{} uses LLM in VFD and VID tasks and extends the region of interest of the detection task from patches to relevant contexts.

\textbf{\textit{LLMs and Vulnerabilities.}} The immense potential of LLMs in code understanding and natural language processing has led to new applications in the field of software security. \cite{steenhoek2024comprehensive} conducted a comprehensive study on the capabilities of LLMs in vulnerability detection, evaluating their performance under various prompting techniques and error types, and comparing their localization capabilities with human developers' performance on real-world vulnerabilities. \cite{li2024enhancing} presented the LLift framework, which enhances practical software vulnerability detection by integrating static analysis with LLMs, particularly in identifying Use Before Initialization (UBI) errors in the Linux kernel. \cite{khare2023understanding} assessed the effectiveness of 16 pre-trained LLMs on 5,000 code samples, covering 25 distinct vulnerability classes in Java and C/C++ languages, and compared them with existing static analysis tools and deep learning-based tools.
These works focus on using LLM to improve vulnerability detection in software or to evaluate the effectiveness of LLMs in vulnerability detection.
\section{Conclusion}
\label{sec:conclusion}
In this paper, we present \tool{}, a tool for identifying vulnerability fixes and vulnerability introduction commits in version management systems using LLM. In VFD, \tool{} analyzes whether a commit has completed the vulnerability fixing work by generating more detailed descriptions and obtaining more relevant information about the patch. In VID, \tool{} traces the history of commits related to vulnerability files, and collects more detailed context of the vulnerability fix patch. The experimental results show that in the VFD task, \tool{}'s precision, recall, and F1-score are better than the baselines, with recall improving by 74\%-77\% compared to SOTA. In the VID task, \tool{}'s precision, recall, and F1-score are also better than the existing SZZ algorithm, and the F1-score is improved by 15\%-27\% compared with other algorithms.

\bibliographystyle{IEEEtranS}
\section*{Acknowledgements}
This work was supported in part by the Key R\&D Program of Hubei Province (2023BAB017, 2023BAB079) and the research funding from MYbank (Ant Group).
\bibliography{main}

\end{document}